\documentclass[a4paper,showpacs,prb,twocolumn]{revtex4}
\usepackage{mathrsfs}
\usepackage{bbm}
\usepackage{amsfonts}
\usepackage{graphicx}
\usepackage{color}
\usepackage{amsmath}
\usepackage{amssymb}
\usepackage{latexsym}
\usepackage{psfrag}
\begin{document}

\title{Topological phases and Majorana states in screened interacting quantum wires}
\author{Hengyi Xu}
\email{hengyi.xu@njnu.edu.cn}
\author{Ye Xiong}
\affiliation{School of Physics and Technology, Nanjing Normal University, Nanjing 210023, China}
\author{Jun Wang}
\affiliation{Department of Physics, Southeast University, Nanjing 210096, China}
\date{\today }

\begin{abstract}
We study theoretically the effects of long-range and on-site Coulomb interactions on the topological phases and transport properties of spin-orbit-coupled quasi-one-dimensional quantum wires imposed on an s-wave superconductor. The electrostatic potential and charge density distributions are computed self-consistently within the Hartree approximation. Due to the finite width of the wires and the charge repulsion, the potential and density distribute inhomogeneously in the transverse direction and tend to accumulate along the lateral edges where the hard-wall confinement is assumed. This result has profound effects on the topological phases and the differential conductance of the interacting quantum wires and their hybrid junctions with superconductors. Coulomb interactions renormalize the chemical potential, and alter the topological phases strongly by enhancing the topological regimes and producing jagged boundaries. Moreover, the multicritical points connecting different topological phases from high-index subbands are modified remarkably in striking contrast to the predictions of the two-band model. We further suggest the possible non-magnetic topological phase transitions manipulated externally with the aid of long-range interactions. Finally, the transport properties of normal-superconductor junctions are also examined and interaction impacts on the emergence of Majorana fermions and the strength of Majorana zero-bias peaks are revealed.  
\end{abstract}

 \maketitle

\section{Introduction}
The existence of Majorana fermions as elementary particles has been a myth since the original proposal suggested by E. Majorana in 1937. \cite{nuocim14.171(1937)} In recent years, condensed matter physicists have been searching for the Majorana fermions as quasi-particle excitations in various solid state hybrid structures with vigorous efforts attributed to some alluring and promising theoretical predictions. \cite{arcmp4.113.(2013),rpp75.076501(2012),prb81.125318(2012)} The enthusiasm was further ignited by the relevant experimental realizations in semiconductor quantum wires with strong spin-orbit couplings and proximity-induced s-wave superconductivity by Mourik {\it et al.}, \cite{science336.1003(2012)} and other groups subsequently \cite{nanolett12.6414(2012),natphy8.887(2012),prl110.126406(2013)}. In these experiments, zero-bias conductance peaks have been observed due to perfect Andreev reflection, signaling the presence of Majorana states at the ends of quantum wires. The experimental measurements show that the zero-bias differential conductance evolves into peaks as the system is tuned into the predicted topological regime without taking into account various effects, such as the finite-length, finite temperature, and electron-electron interactions etc. To clarify some discrepancies between experiments and theories, the effects of disorder \cite{prl109.267002(2012),njp14.125011(2012)}, nonclosure of gaps \cite{prl109.266402(2012)}, inhomogeneous pairing potentials \cite{prl110.186803(2013)} have been investigated theoretically. More severely, alternative mechanisms, for examples, the Andreev bound state \cite{prb65.184505(2002)} and the Kondo effect \cite{nat405.764(2000)} which also produce the zero-bias peaks have been suggested to challenge the experimental findings. 

Among all the aforementioned effects, the electronic interaction is of vital importance and tricky to treat microscopically. \cite{prb88.161103(R)(2013)} It is expected that Coulomb interactions can strongly influence the stability of Majorana modes \cite{prl107.036801(2011),prb84.085114(2011),njp14.125018(2012)}, and are, therefore, crucial for understanding quantitatively the experimental findings and ultimately recognition of the existence of Majorana bound states at the ends of quantum wires. In the one-dimensional (1D) quantum wires, repulsive interacting electrons form interacting Luttinger liquids and should be described more precisely by the corresponding theory. \cite{prb85.245121(2012),prb84.085114(2011)} To attack this problem, various methods have been employed. Based on the density matrix renormalization group (DMRG), tunneling spectra of interacting Kitaev chains and Majorana edge states have been examined. \cite{prb88.161103(R)(2013)} In particular, E. Stoudenmire {\it et al.}, \cite{prb84.014503(2011)} compared systematically the DMRG, Hartree-Fock, and bosonization approaches for treating the interacting Majorana wires and found that the interaction problem can be described reasonably well using Hartree-Fock theory with the sufficiently strong proximity effect and applied magnetic fields albeit it deserves more powerful DMRG and bosonization techniques. Besides the single-mode wires, multichannel wires have also been studied considerably. \cite{prl105.227003(2010),prb84.214528(2011),prl105.046803(2010),prb83.094525(2011),prl106.127001(2011)}  Lutchyn {\it et al.}, \cite{prb84.214528(2011)} have studied the roles of interactions on the low-energy topological phase diagram near the multicritical point connecting the topological phases originating from the first and second transverse subbands, and revealed that the interactions renormalize the phase boundary near the multicritical point leading to the hybridization of Majorana modes from different subbands. Furthermore, the presence of disorder was found to induce the phase transition from topological phases to trivial localized phases together with interactions. \cite{prl109.146403(2012)}  

In a realistic experimental setup, the semiconducting quantum wire with a high g-factor and spin-orbit coupling is exposed on a metallic s-wave superconductor to get a proximity energy gap. The metallic superconductor, as a secondary effect, may drive the electronic interactions into a strongly screened regime. \cite{jpc26.172203(2014)} Consequently, the electronic density and potential distributions in multiband nanowires are rather inhomogeneous along the transverse direction due to the finite width and electronic repulsions. This inhomogeneity in electrostatic potential can be one of the major sources of the soft superconducting gaps. For the transport properties of the semiconductor-superconductor hybrid structures, much of the prior work has been focusing on the non-interacting cases. \cite{prb91.024514(2015),prb91.2145413(2015),prb85.245121(2012),prb88.064509(2013),njp15.075019(2013),prb90.115107(2014)} How the screened interactions and the inhomogeneous potential distribution influence the topological phases in multiband quantum wires and the related Majorana modes, has received relatively less attention. In this work, we study the topological phases and Majorana zero mode in a typical experiment-relevant semiconductor-superconductor hybrid device composed of an interacting quantum wire in proximity to an s-wave superconductor. The screened Coulomb interactions are incorporated by the self-consistent Hartree-Fock calculations in the presence of external magnetic fields. It is shown that electron-electron interactions strongly change the energy bands and modify the topological phase boundaries as well as the emergence of Majorana modes. 

The paper is organized as follows. In Sec. II we introduce the structure to be investigated and formulate our model. The calculation results are presented and discussed in Sec. III. Sec. IV contains the summary and conclusions.

\section{Theoretical Model}
We consider a spin-orbit-coupled semiconductor quantum wire of the width $W$ in the y-direction and the length $L$ along the x-direction deposited on an s-wave superconducting electrode, while its left side is contacted through a tunnel barrier $U_p$ by a normal metallic lead as shown in Fig. \ref{fig1}(a). The s-wave superconductor induces a paring potential $\Delta$ for the electrons in the wire.  The whole system is subjected to a uniform in-plane magnetic field $B_x$. Throughout the calculations, we choose the realistic parameters for InSb semiconductor quantum wires: $\Delta=0.25\mathrm{meV}$, g-factor $\mathrm{g}=50$, Rashba spin-orbit coupling strength $t_R=20 \mathrm{meV\cdot nm}$, and effective mass $m^*=0.015m_e$ with $m_e$ being the electron mass.

The system is described by the tight-binding Hamiltonian consisting of three terms as 
\begin{equation}
\mathcal{H}=H_0+H_R+H_U,\label{hamtt}
\end{equation}
with respective form given by 
\begin{eqnarray}
H_0&=&\sum_{i,\sigma} c_{i,\sigma}^\dag (\epsilon_{0}+V_H-\mu)c_{i,\sigma}-t\sum_{\langle i,j\rangle,\sigma} c_{i,\sigma}^\dag c_{j,\sigma} \nonumber\\
&&+\frac{1}{2}g\mu_B \sum_{i;\sigma,\sigma'}c_{i,\sigma}^\dag s_xB_x c_{i,\sigma'};\label{hamt0}
\end{eqnarray}
\begin{equation}
H_R=it_R\sum_{\langle i,j\rangle;\sigma,\sigma'} {\mathbf {\hat e}_z}\cdot (\mathbf s \times \mathbf d_{ij})c_{i,\sigma}^\dag c_{j,\sigma'};\label{hamtr}
\end{equation}
\begin{equation}
H_U=U\sum_{i,\sigma}n_{i\sigma}n_{i\bar\sigma}.\label{hamtu}
\end{equation}
where $c_{i,\sigma}^\dag$ and $c_{i,\sigma}$ are creation and annihilation operators for an electron with spin $\sigma (\uparrow,\downarrow)$ on site $i$, and $\mathbf s$ denotes the Pauli matrices. 
The Hamiltonian $H_0$ in Eq. (\ref{hamt0}) represents the Hamiltonian of semiconductor quantum wires including the on-site energy $\epsilon_0=-4t$ and the hopping energy $t$ between the nearest-neighbouring sites along $x$- and $y$-directions. $V_H$ and $\mu$ are the Hartree potential and the chemical potential, respectively, and the last contribution is from the Zeeman splitting due to the in-plane magnetic field $B_x$. The term $H_R$ in Eq. (\ref{hamtr}) describes the Rashba spin-orbit coupling with $\mathbf d_{ij}$ being a lattice vector pointing from site $j$ to site $i$. $\langle\rangle$ runs over all the nearest-neighbouring sites. The on-site electronic interactions between electrons of the opposite spins are captured by the Hubbard-like term $H_U$ in Eq. (\ref{hamtu}). To facilitate the computation, Eq. (\ref{hamtu}) can be rewritten within the mean-field approximation such that a charge with spin $\sigma$ at the site $\mathbf r_i$ interacts with the average charge population with an opposite spin $\langle n_{\bar \sigma}\rangle$  at the same site and vice versa. Moreover, the Hartree term $V_H(\mathbf r)$ in Eq. (\ref{hamt0}) depicts the long-range Coulomb interactions between charges at different sites in the semiconducting quantum wire, \cite{prb73.075331(2006),jpc26.172203(2014)}
\begin{equation}
V_H(\mathbf r_{i})=\frac{e^2}{4\pi\epsilon_0\epsilon_r}\sum_{\mathbf r_i\neq \mathbf r_j}n(\mathbf r_j)\left(\frac{1}{|\mathbf r_i-\mathbf r_j|}-\frac{1}{\sqrt{|\mathbf r_i-\mathbf r_j|^2+4d^2}} \right).\label{vhart}
\end{equation}
where $d$ is the distance between the quantum wire and the superconducting metallic gate, and the second part in the parenthesis is the contribution from the mirror charges due to the presence of the metallic superconducting gate. The average charge population at the site $\mathbf r_i$ is calculated by 
\begin{equation}
\langle n_{\sigma}(\mathbf r_i)\rangle =-\frac{1}{\pi}\int_{-\infty}^{E_F}\Im[G_\sigma(\mathbf r_i,\mathbf r_i;E)] f_{FD}(E,E_F)dE,\label{nrho}
\end{equation} 
where $G_\sigma(\mathbf{r}_i,\mathbf{r}_i;E)$ is the Green's function on the site $\mathbf r_i$ at energy $E$ for spin $\sigma$. From the computational point of view, both short- or long-range Coulomb interactions affect only the diagonal elements of the Hamiltonian matrix. Eqs. (\ref{hamtt})-(\ref{nrho}) can be solved self-consistently starting from some initial guess of charge density $\langle n_{\sigma}\rangle$ to obtain the self-consistent charge and electrostatic potential distributions which can be further used to calculate the band structures, the interacting topological phase diagrams and the differential conductance spectroscopy.   

We study the topological phase diagram by calculating a $Z_2$ analogy topological invariant in quasi-1D wires, namely the evolution of Wannier function center during a pumping process, which has been formulated in detail in Ref. [\onlinecite{prb84.075119(2011)}]. Here we only give a very brief description of this method. The main idea of this formalism is to investigate the maximally localized evolution of Wannier functions for quasi-1D systems by studying the eigenstates of the position operator projected into the subspace of the occupied states. In the eigenstate space, the projected position operator can be written in a matrix form with nonzero super-diagonal and left-down corner elements. The nonzero elements of the matrix are again the matrices formed by the products of all occupied eigenvectors. The eigenproblem of the block position operator can be solved by constructing the matrix $D(k_y)$, a product of all of its nonzero block matrices. Equivalently, $D(k_y)$ can be viewed as a product of the Berry connections along the so-called ``Wilson loop'' and further expressed in the language of non-Abelian gauge fields $A^{mn}_{i,i+1}$ as $D(k_y)=\Pi_{i=0}^{N_x-1} e^{-iA_{i,i+1}\delta k}$ with $N_x$ being the number of discrete $k_x$ and $\delta k=k_x^{i+1}-k_x^i$. The phase factor $\theta^D_m$ of the $m$-th eigenvalue of $D(k_y)$ determines the evolution of the Wannier function center pairs which reside in a cylinder surface and enclose it integer times for an effective 1D system with $k_y$. The enclosing times equivalent to the winding number of the Wannier center pairs are used to distinguish the different phases in our study.

The calculations are started by performing the Fourier transformation along $x$-direction since $k_x$ is a good quantum number, while the Hamiltonian in the $y$-direction remains in the real-space.  Thus, the Hamiltonian for a discretized site $i$ in the momentum space is given by
\begin{equation}
\left[\begin{array}{cccc}
h_\uparrow & 2it_R\sin(k_x) &  & \Delta \\ 
-2it_R\sin(k_x) & h_\downarrow & -\Delta  &  \\ 
 & -\Delta^* & -h_\uparrow^* & -2it_R\sin(k_x) \\ 
\Delta^＊ &  & 2it_R\sin(k_x) & -h_\downarrow^*
\end{array}  \right],\label{hamtkr}
\end{equation}
with $h_\sigma(\mathbf r_i)=\epsilon_0-2t\cos(k_x)-\mu +V_H(\mathbf r_i)+V_U(\mathbf r_i)\pm E_z$ and $\sigma=(\uparrow,\downarrow)$, and $V_U$ is the potential from Eq. (\ref{hamtu}). The nearest-neighboring sites along the y-direction are coupled by the matrix 
\begin{equation}
\left[\begin{array}{cccc}
t & -it_R/2 &  &  \\ 
it_R/a & t &  &  \\ 
 &  & -t & -it_R/2 \\ 
 &  & -it_R/2 & -t
\end{array}  \right]. \label{hamtcp} 
\end{equation}
The total Hamiltonian $\mathcal H$ can be written in the block matrix form with the diagonal  part in the form of Eq. (\ref{hamtkr}) and superdiagonal or subdiagonal part in the form of Eq. (\ref{hamtcp}).
By directly diagonalizing $\mathcal H$ in the momentum-real mixed space, one can obtain the wave functions of the eigenstates and furthermore the matrix $D(k_y)$. For interacting cases, the self-consistent Hartree and Hubbard potentials are used in Eq. (\ref{hamtkr}). 

The differential conductance of normal-superconductor (NS) junctions is calculated by 
\begin{equation}
\frac{dI}{dV}=\frac{e^2}{h}\left[N-R_{ee}+R_{eh} \right],
\end{equation}
where $N$ is the number of propagating modes in the normal lead, and $R_{ee}$ and $R_{eh}$ are the normal and Andreev reflections, respectively. The calculations of the reflections are based on the Nambu Green's function technique which has been formulated in detail in Ref. [\onlinecite{prb77.245401(2008),pla377.3148(2013)}].

\section{Results and discussion}
\begin{figure}[tbp]
\includegraphics[scale=1]{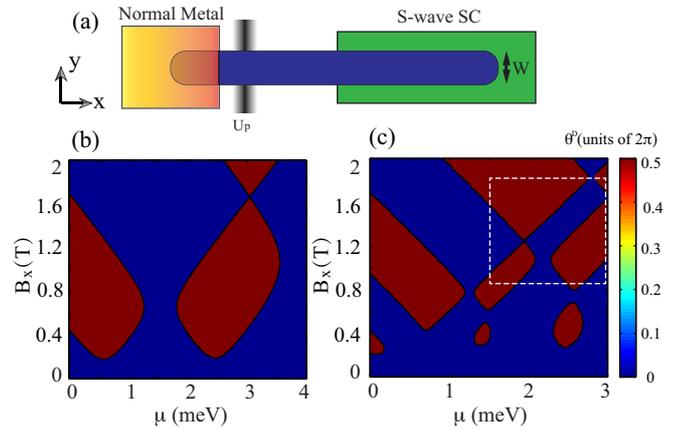} 
\caption{(Color online) (a) The schematic structure of the normal metal-superconductor (NS) junction for the transport properties calculations. Lower panels: the phase diagrams as a function of the chemical potential and the magnetic field for width (b) $W=200\mathrm{nm}$ and (c) $W=400\mathrm{nm}$. The white rectangular frame in (c) indicates the regime to be considered in the interacting case. Other parameters are $\Delta=0.25\mathrm{meV}$, $g=50$, $t_R=20\mathrm{meV\cdot nm}$, and $m^*=0.015m_e$.}
\label{fig1}
\end{figure}
As the first step, we study the non-interacting phase diagrams in wide parameter ranges by calculating the phase factor $\theta^D_m$ associated with the evolution of the Wannier function center for quasi-1D wires. Fig. \ref{fig1}(b) and (c) show the topological invariants as a function of the chemical potential $\mu$ and the applied magnetic field $B_x$ for different wire widths. As the parameters vary, two phase factors $\theta^D_m$ split and meet again resulting in an integer times of $2\pi$ difference in their values, which is equivalent to the winding number of Wannier center pairs. For $W=200\mathrm{nm}$ (see Fig. \ref{fig1}(a)), the multichannel quantum wire exhibits three topological phase regimes separated by topological trivial regimes in the calculated parameter range. The topological phases with a value of $0.5$ in units of $2\pi$ appear as rounded rectangular blocks originating from different subbands in the phase diagram. As the wire width increases, more topological regimes associated with high-energy subbands show up as shown in Fig. \ref{fig1}(c) since the energy separations between subbands decrease with the wire width.

\begin{figure}[tbp]
\includegraphics[scale=1]{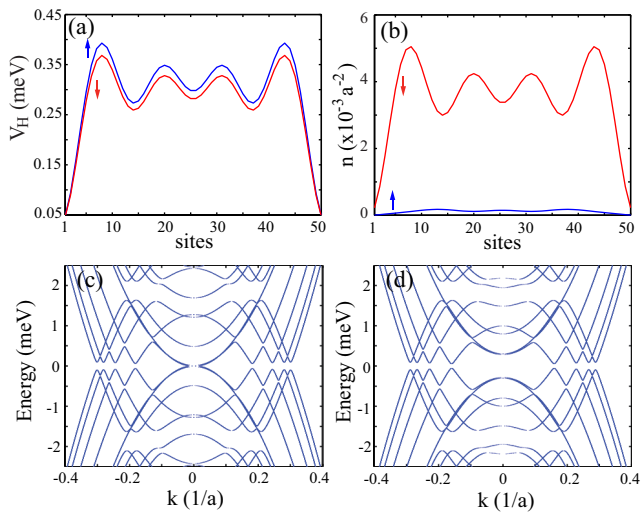} 
\caption{(Color online) The self-consistent (a) Hartree potential together with Hubbard term and (b) charge density profiles at the multi-critical point with $\mu=1.9\mathrm{meV}$ and $B_x=1.2\mathrm{T}$. Comparison between the non-interacting (c) and interacting (d) band structures at the multicritical point. $\varepsilon_r=18$ for $\mathrm{InSb}$ and $d=10\mathrm{nm}$.}
\label{fig2}
\end{figure}

For both cases, the topological regimes along the direction of sweeping $\mu$ with fixed $B_x$ are disconnected. Their separations are basically the distance between the neighboring subbands with different spins. In contrast, the topological regimes may contact each other forming the so-called multicritical point with varying the magnetic field at a fixed chemical potential. The emergence of multicritical points owes to that two subbands are very close and even touch each other at low energies around $k=0$ such that two topological phase transitions occur continuously and even simultaneously, see Fig.\ref{fig2}(c). Moreover, relatively small subband separations of the wider wire give rise to small topological islands at low magnetic fields corresponding to the first subband as shown in Fig. \ref{fig1}(c). It should be noted that the noninteracting phase diagrams of quasi-1D nanowires have also been examined and a similar phase diagram was found. \cite{prb86.024505(2012)}

When Coulomb interactions are turned on, the charges distribute inhomogeneously along the transverse direction due to long-range repulsive potentials computed by Eq. (\ref{vhart}) as shown in Fig. \ref{fig2}. We first focus on the effects of Coulomb interactions on the multicritical point in the white frame shown in Fig. \ref{fig1}(c). Fig. \ref{fig2}(a) and (b) show the electronic potential and corresponding charge density profiles in the transverse direction at the critical point where one topological regime transits to another. The application of gate voltages can tune the chemical potential of the system and vary the charge density. The overall charge distribution shows a density enhancement toward the wire edges, which is related to the effect of the electrostatic Coulomb repulsion in a hard-wall confined structure. The applied magnetic field lifts the spin degeneracy, while the short-range Coulomb potential accounted by Eq. (\ref{hamtu}) further enhances the Zeeman splitting leading to an asymmetry between the spin-up and spin-down branches in the potential. As a result, the opposite spin subbands are not equally populated as indicated in charge density profile (see Fig. \ref{fig2}(b)). The roles of on-site Coulomb interactions in single-mode wires have been studied previously by Stoudenmire {\it et al.,} \cite{prb84.014503(2011)} and the conclusions are equally applied here for multichannel nanowires. The experimental significance of on-site interactions is to lower the critical magnetic field to enter the topological phase with Majorana modes existing at two ends of the wire.

\begin{figure}[tbp]
\includegraphics[scale=1]{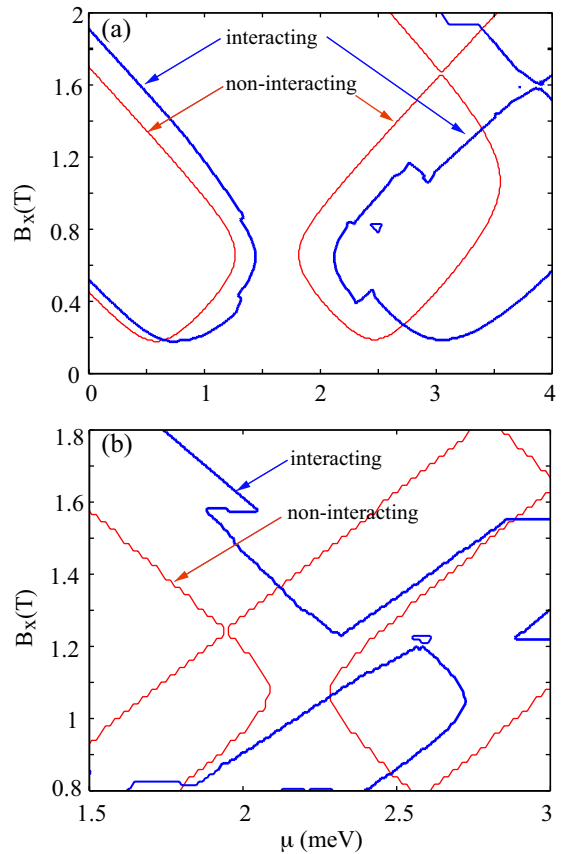} 
\caption{(Color online) The interacting phase diagrams as a function of the magnetic field $B_x$ and the chemical potential $\mu$ for (a) $W=200\mathrm{nm}$ and (b) $W=400\mathrm{nm}$. The non-interacting phase diagrams (Red thin) have been included for comparisons. The parameters range in (b) are indicated by the white rectangular frame in Fig. \ref{fig1}(c).}
\label{fig3}
\end{figure}

Fig. \ref{fig2}(c) shows the corresponding non-interacting band structure of the multicritical point at $\mu=1.9\mathrm{meV}$ and $B_x=1.2\mathrm{T}$ where the 3rd-subband spin-up branch overlaps with the lower index subbands. As the magnetic field increases, the particle and hole subbands approach each other and a topological phase transition results with a process of the gap closure and reopening. The system may undergo a series of phase transition processes as the gap is further closed and reopened by the subsequent subbands. In the presence of Coulomb interactions as shown in Fig. \ref{fig2}(d), the low-energy subbands around small wave vectors retreat towards particle and hole directions, respectively, while the minigap where particle and hole subbands anticross remains unchanged. Evidently, the interactions do not spoil the overlap of two subbands around the zero energy but defer the emergence of the multicritical point. It will be clear in the following that the influences of interactions on mutlitcritical points may vary according to their subband originations. 

In Fig. \ref{fig3}, we show the roles of Coulomb interactions in different quantum wires with broad parameter ranges. The inhomogeneous electrostatic potential renormalizes the chemical potential due to electronic repulsions and shifts the topological phases to high chemical potentials. Also, the interactions enlarge the areas of topological regimes, which is consistent with the single-mode cases where on-site interactions increase the chemical potential windows for fixed magnetic fields, thereby enhancing the system immunity against the fluctuation of $\mu$. \cite{prb84.014503(2011)} Most strikingly, the long-range Coulomb interactions strongly modify the boundaries of topological phases and even generate jagged boundaries. It is also evident that some very small isolate topological regimes may be produced near the phase boundaries, and topologically trivial phases may appear in topological areas as well. This is because, close to the phase boundaries, small perturbations are sufficient to introduce or interrupt a closure of the energy gap so that additional phase transitions occur. For the case of $W=400\mathrm{nm}$, the topological island at $B_x=1.2\mathrm{T}$ and $\mu=2.6\mathrm{meV}$ appears as a remnant of the non-interacting multicritical point since the subband overlap is preserved as shown in Fig. \ref{fig2}(d). It has been pointed out the repulsive on-site interactions affect only quantitatively the topological phases and the appearance of multicritical points remains similar. \cite{prb84.214528(2011)} In our cases, the impact of long-range interactions on the multicritical point in the narrow wire is predominantly a shift to higher $\mu$ in agreement with the two-band model in Ref. [\onlinecite{prb84.214528(2011)}]. By contrast, long-range Coulomb interactions influence the multicritical points formed by the high-index subbands in a more profound way. From Fig. \ref{fig3}(b), one can see clearly that the two phase boundary vertexes do not coincide with each other in chemical potential any longer due to interactions. 
Thus, the multicritical points from high-index subbands are more fragile to the inhomogeneity of Coulomb potentials along the transverse direction.

In fact, Coulomb interactions may play a much more crucial role in the manipulation of topological phase transitions. In graphene, San-Jose {\it at al.}, \cite{prx5.041042(2015)} has recently put forward a Majorana zero-mode mechanism through interactions without the aid of spin-orbit couplings. This is a very significant progress in this field and pave a way for constructing graphene Majorana in view of the rather weak spin-orbit coupling of graphene. For long-range Coulomb interactions, their major advantage in the present case is that they can be tuned externally by gate voltages. Therefore, the long-range interaction can be an important tool at hand which enables ones to manipulate the phase transitions conveniently. It has been shown that long-range Coulomb interactions can generate band-structure warping and lead to an anomalous conductance reduction in graphene indicating a possible topological transition. \cite{prb82.115311(2010),prb79.035421(2009)} Based on these considerations, study of effects of interactions in nanowires are extremely important for its promising realization of a non-magnetic topological phase transition and the related Majorana quasi-particle.

\begin{figure}[tbp]
\includegraphics[scale=1]{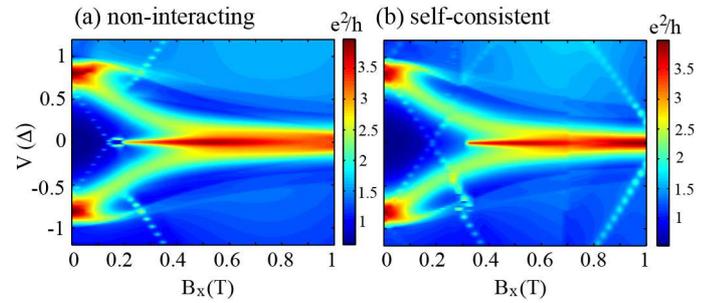} 
\caption{The differential conductance as a function of the bias voltage and the magnetic field for the cases of the (a) non-interacting and (b) interacting quantum wires. The system has parameters $\mu=2.5\mathrm{meV}$, $W=200\mathrm{nm}$ and $L=200\mathrm{nm}$. The pinch-off gate $U_p=8\mathrm{meV}$ with width $d=10\mathrm{nm}$.}
\label{fig4}
\end{figure}

\begin{figure}[tbp]
\includegraphics[scale=1]{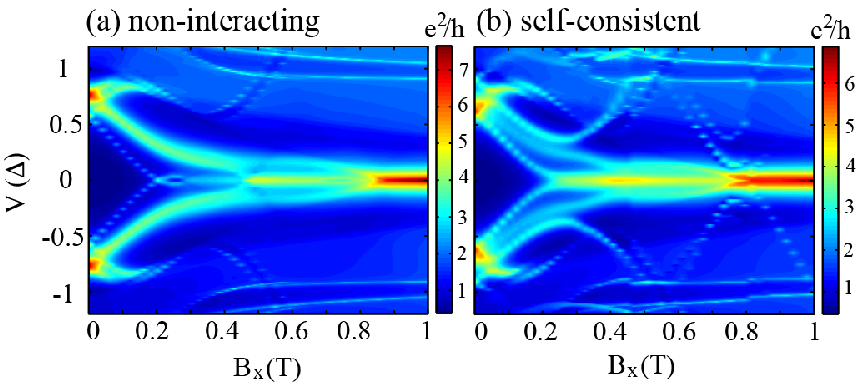} 
\caption{The differential conductance as a function of the bias voltage and the magnetic field for the cases of the (a) non-interacting and (b) interacting quantum wires. The system has parameters $\mu=1.4\mathrm{meV}$, $W=400\mathrm{nm}$ and $L=200\mathrm{nm}$. The pinch-off gate $U_p=8\mathrm{meV}$ with width $d=10\mathrm{nm}$.}
\label{fig5}
\end{figure}

\begin{figure}[tbp]
\includegraphics[scale=1]{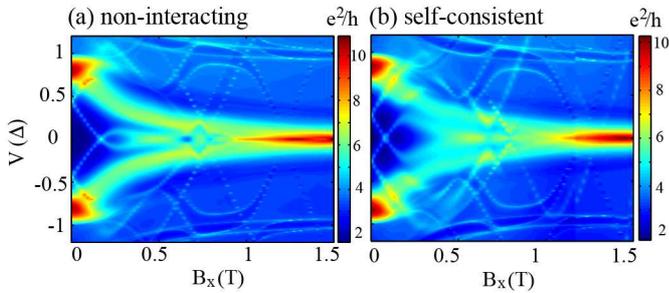} 
\caption{The differential conductance as a function of bias voltage and magnetic field for the cases of the (a) non-interacting and (b) interacting quantum wires. The system has parameters $\mu=2.8\mathrm{meV}$, $W=400\mathrm{nm}$ and $L=200\mathrm{nm}$. The pinch-off gate $U_p=8\mathrm{meV}$ with width $d=10\mathrm{nm}$.}
\label{fig6}
\end{figure}

We proceed by studying the differential conductance $dI/dV$ of quantum wires with a superconducting metal as the right lead. In our case, we consider the simplest hybrid structure, namely a normal-superconductor junction as shown in Fig. \ref{fig1}(a). Various hybrid junctions hosting Majorana fermions have been studied extensively, and a rich phenomenology has been revealed. \cite{prb86.180503(R)(2012)} Here we primarily concern with the interaction effects on the differential conductance spectroscopy such that a simple geometry is more illuminating. Fig. \ref{fig4}-\ref{fig6} show the differential conductance as a function of the bias voltage $V$ and the magnetic field $B_x$ for different widths and chemical potentials. The left and right panels correspond to the non-interacting and interacting cases, respectively. 

In Fig. \ref{fig4}, the non-interacting $dI/dV$ peaks around the gap edges at zero magnetic field. With the increase of the magnetic field, the gap slowly shrinks and closes at the critical point of phase transitions forming a prototypical Y-type structure in the $B_x-V$ (magnetic field-bias voltage) plane. As the gap is closed and reopened, the zero-bias peak (ZBP) develops implying the formation of Majorana end states. Fig. \ref{fig4}(b) shows the interacting differential conductance through the topological superconductor with the self-consistent potential which is calculated for different magnetic fields. The commencement of the ZBP moves towards higher magnetic fields and the peak strength becomes even stronger at high magnetic fields compared with the non-interacting case. This change in the ZBP can be traced back to the phase diagram in Fig. \ref{fig3}(a). In the presence of Coulomb interactions, the phase boundary at $\mu=2.5\mathrm{meV}$ is displaced with respect to the magnetic field compared with non-interacting case consistent with the changes in the ZBP quantitatively. However, this consistency is just a serendipity because the phase diagrams are obtained for infinitely long wires, while the $dI/dV$ spectroscopy is calculated through semi-infinite topological superconductors. For wider wires, their discrepancies are evident. 

Fig. \ref{fig5} shows the wider quantum wire with $W=400\mathrm{nm}$ and a smaller chemical potential $\mu=1.4\mathrm{meV}$. Besides the similar Y-type structure, the $dI/dV$ spectroscopy exhibits richer structures. The first visible peak appears at $B_x=0.2\mathrm{T}$ for the non-interacting wire, which is due to the Andreev bound states since it splits with the magnetic field. The ZBP induced by Majorana zero-modes arises at $B_x=0.3\mathrm{T}$ and builds up as the magnetic field increases. However, the ZBP is interrupted with the increasing of magnetic fields by the phase transitions originating from different subbands. This can be verified by the phase diagram shown in Fig. \ref{fig1}(c). Fig. \ref{fig5}(b) shows the effects of Coulomb interactions on differential conductance spectroscopy. The resonance from Andreev bound states at $\mu=0.2\mathrm{meV}$ is diminished and the ZBP from Majorana modes becomes stronger and more recognizable in intermediate magnetic fields. We further consider the wire with higher chemical potential as shown in Fig. \ref{fig6}. Compared with the non-interacting case, the interactions modifies the particle and hole gap edges reducing the minigap which is reflected resonances within the bulk gap at the zero magnetic field in Fig. \ref{fig6}(b). The ZBP around $B_x\approx 1\mathrm{T}$ from Majorana mode is suppressed accompanied with the modification of phase transitions due to Coulomb interactions. In addition, the interactions also influence the position of resonances due to Andreev bound states. 

From above transport calculations, we arrive at some points on Majorana fermion observations. (i) long-range Coulomb interactions modify the topological phase boundary and affect the position of ZBPS in the parameter space. In experiments, the Majorana ZBP may appear at quite different places with the non-interacting theoretical predictions for multichannel wires.  
This will no doubt cause the difficulties in the recognition of Majorana fermions. (ii) Coulomb interaction can change topological windows and number of phase transitions thereby strongly altering the appearance of Majorana ZBPs. Theoretical predictions based on single-particle models is unreliable for interpreting the experimental findings particularly for high chemical potential in multiband wires.

We end this section with the following comments. Throughout our calculations, we fixed the pinch-off gate voltage and wire length. It is expected that pinch-off gates and wire length only affect the resonances in trivial phases, and have no significant effects on topological regimes. Also, our calculations are performed at zero temperature. We expect that finite temperature will influence the results quantitatively not qualitatively. Our discussion and conclusions remain valid. Finally, the interplay between Coulomb interactions and disorder is an interesting question. Most recent researches concentrate on the disorder effects in the normal region of NS junctions. How disorder influences the topological phases together with the long-range Coulomb interaction is desirable for both experimental purposes and theoretical interest. However, this issue is beyond the scope of the present work and will be the central task in our further work.

\section{Summary and Conclusions}
We have calculated the phase diagrams of an infinitely long quasi-1D wire with a high spin-orbit interaction imposed on an s-wave superconductor. The phase diagrams in $B_x-\mu$ plane demonstrate connected or disconnected topological phase regimes originating from different subbands. We incorporate Coulomb interactions in a self-consistent way within the Hartree approximation and study the potential and charge density distributions along the transverse direction. The repelling charges tend to accumulate along the wire edges and form the characteristic oscillations in profile. The inhomogeneously distributed potentials alter the band structures strongly and modify the phase diagrams in a nontrivial way. In particular, the multicritical points from high-index subbands respond to the interactions more prominently compared with those from low-index subbands.

The changes in the phase diagram due to interactions can be experimentally detected by transport measurements, namely $dI/dV$ spectroscopy. The study of the roles of interactions in $dI/dV$ is particularly important for recognition of Majorana fermions in experiments. For multiband wires, $dI/dV$ spectroscopy with interactions may be dramatically different compared with the non-interacting cases. The careful investigation of the effects of Coulomb interactions on the transport properties has a far-reaching significance for identification of Majorana fermions in experiments.

\begin{acknowledgments}
H.X. acknowledges financial support from Department of Education of Jiangsu province through Grant No. 164080H00210. J.W. thanks the support from NSFC (Grant No. 115074045).
\end{acknowledgments}


\begin{thebibliography}{41}
\expandafter\ifx\csname natexlab\endcsname\relax\def\natexlab#1{#1}\fi
\expandafter\ifx\csname bibnamefont\endcsname\relax
  \def\bibnamefont#1{#1}\fi
\expandafter\ifx\csname bibfnamefont\endcsname\relax
  \def\bibfnamefont#1{#1}\fi
\expandafter\ifx\csname citenamefont\endcsname\relax
  \def\citenamefont#1{#1}\fi
\expandafter\ifx\csname url\endcsname\relax
  \def\url#1{\texttt{#1}}\fi
\expandafter\ifx\csname urlprefix\endcsname\relax\def\urlprefix{URL }\fi
\providecommand{\bibinfo}[2]{#2}
\providecommand{\eprint}[2][]{\url{#2}}

\bibitem[{\citenamefont{Majorana}(1937)}]{nuocim14.171(1937)}
\bibinfo{author}{\bibfnamefont{E.}~\bibnamefont{Majorana}},
  \bibinfo{journal}{Il Nuo. Cim.} \textbf{\bibinfo{volume}{14}},
  \bibinfo{pages}{171} (\bibinfo{year}{1937}).

\bibitem[{\citenamefont{Beenakker}(2013)}]{arcmp4.113.(2013)}
\bibinfo{author}{\bibfnamefont{C.~W.~J.} \bibnamefont{Beenakker}},
  \bibinfo{journal}{Annu. Rev. Con. Mat. Phys.} \textbf{\bibinfo{volume}{4}},
  \bibinfo{pages}{113} (\bibinfo{year}{2013}).

\bibitem[{\citenamefont{Alicea}(2012)}]{rpp75.076501(2012)}
\bibinfo{author}{\bibfnamefont{J.}~\bibnamefont{Alicea}},
  \bibinfo{journal}{Rep. Prog. Phys.} \textbf{\bibinfo{volume}{75}},
  \bibinfo{pages}{076501} (\bibinfo{year}{2012}).

\bibitem[{\citenamefont{Alicea}(2010)}]{prb81.125318(2012)}
\bibinfo{author}{\bibfnamefont{J.}~\bibnamefont{Alicea}},
  \bibinfo{journal}{Phys. Rev. B} \textbf{\bibinfo{volume}{81}},
  \bibinfo{pages}{125318} (\bibinfo{year}{2010}).

\bibitem[{\citenamefont{Mourik et~al.}(2012)\citenamefont{Mourik, Zuo, Frolov,
  Plissard, Bakkers, and Kouwenhoven}}]{science336.1003(2012)}
\bibinfo{author}{\bibfnamefont{V.}~\bibnamefont{Mourik}},
  \bibinfo{author}{\bibfnamefont{K.}~\bibnamefont{Zuo}},
  \bibinfo{author}{\bibfnamefont{S.}~\bibnamefont{Frolov}},
  \bibinfo{author}{\bibfnamefont{S.}~\bibnamefont{Plissard}},
  \bibinfo{author}{\bibfnamefont{E.}~\bibnamefont{Bakkers}}, \bibnamefont{and}
  \bibinfo{author}{\bibfnamefont{L.}~\bibnamefont{Kouwenhoven}},
  \bibinfo{journal}{Science} \textbf{\bibinfo{volume}{336}},
  \bibinfo{pages}{1003} (\bibinfo{year}{2012}).

\bibitem[{\citenamefont{Deng et~al.}(2012)\citenamefont{Deng, Yu, Huang,
  Larsson, Caroff, and Xu}}]{nanolett12.6414(2012)}
\bibinfo{author}{\bibfnamefont{M.}~\bibnamefont{Deng}},
  \bibinfo{author}{\bibfnamefont{C.}~\bibnamefont{Yu}},
  \bibinfo{author}{\bibfnamefont{G.}~\bibnamefont{Huang}},
  \bibinfo{author}{\bibfnamefont{M.}~\bibnamefont{Larsson}},
  \bibinfo{author}{\bibfnamefont{P.}~\bibnamefont{Caroff}}, \bibnamefont{and}
  \bibinfo{author}{\bibfnamefont{H.}~\bibnamefont{Xu}}, \bibinfo{journal}{Nano
  Lett.} \textbf{\bibinfo{volume}{12}}, \bibinfo{pages}{6414}
  (\bibinfo{year}{2012}).

\bibitem[{\citenamefont{Das et~al.}(2012)\citenamefont{Das, Ronen, Most, Oreg,
  Heiblum, and Shtrikman}}]{natphy8.887(2012)}
\bibinfo{author}{\bibfnamefont{A.}~\bibnamefont{Das}},
  \bibinfo{author}{\bibfnamefont{Y.}~\bibnamefont{Ronen}},
  \bibinfo{author}{\bibfnamefont{Y.}~\bibnamefont{Most}},
  \bibinfo{author}{\bibfnamefont{Y.}~\bibnamefont{Oreg}},
  \bibinfo{author}{\bibfnamefont{M.}~\bibnamefont{Heiblum}}, \bibnamefont{and}
  \bibinfo{author}{\bibfnamefont{H.}~\bibnamefont{Shtrikman}},
  \bibinfo{journal}{Nature Phys.} \textbf{\bibinfo{volume}{8}},
  \bibinfo{pages}{887} (\bibinfo{year}{2012}).

\bibitem[{\citenamefont{Finck et~al.}(2013)\citenamefont{Finck, Van~Harlingen,
  Mohseni, Jung, and Li}}]{prl110.126406(2013)}
\bibinfo{author}{\bibfnamefont{A.}~\bibnamefont{Finck}},
  \bibinfo{author}{\bibfnamefont{D.}~\bibnamefont{Van~Harlingen}},
  \bibinfo{author}{\bibfnamefont{P.}~\bibnamefont{Mohseni}},
  \bibinfo{author}{\bibfnamefont{K.}~\bibnamefont{Jung}}, \bibnamefont{and}
  \bibinfo{author}{\bibfnamefont{X.}~\bibnamefont{Li}}, \bibinfo{journal}{Phys.
  Rev. Lett.} \textbf{\bibinfo{volume}{110}}, \bibinfo{pages}{126406}
  (\bibinfo{year}{2013}).

\bibitem[{\citenamefont{Liu et~al.}(2012)\citenamefont{Liu, Potter, Law, and
  Lee}}]{prl109.267002(2012)}
\bibinfo{author}{\bibfnamefont{J.}~\bibnamefont{Liu}},
  \bibinfo{author}{\bibfnamefont{A.~C.} \bibnamefont{Potter}},
  \bibinfo{author}{\bibfnamefont{K.}~\bibnamefont{Law}}, \bibnamefont{and}
  \bibinfo{author}{\bibfnamefont{P.~A.} \bibnamefont{Lee}},
  \bibinfo{journal}{Phys. Rev. Lett.} \textbf{\bibinfo{volume}{109}},
  \bibinfo{pages}{267002} (\bibinfo{year}{2012}).

\bibitem[{\citenamefont{Pikulin et~al.}(2012)\citenamefont{Pikulin, Dahlhaus,
  Wimmer, Schomerus, and Beenakker}}]{njp14.125011(2012)}
\bibinfo{author}{\bibfnamefont{D.}~\bibnamefont{Pikulin}},
  \bibinfo{author}{\bibfnamefont{J.}~\bibnamefont{Dahlhaus}},
  \bibinfo{author}{\bibfnamefont{M.}~\bibnamefont{Wimmer}},
  \bibinfo{author}{\bibfnamefont{H.}~\bibnamefont{Schomerus}},
  \bibnamefont{and}
  \bibinfo{author}{\bibfnamefont{C.}~\bibnamefont{Beenakker}},
  \bibinfo{journal}{New J. Phys.} \textbf{\bibinfo{volume}{14}},
  \bibinfo{pages}{125011} (\bibinfo{year}{2012}).

\bibitem[{\citenamefont{Stanescu et~al.}(2012)\citenamefont{Stanescu, Tewari,
  Sau, and Sarma}}]{prl109.266402(2012)}
\bibinfo{author}{\bibfnamefont{T.~D.} \bibnamefont{Stanescu}},
  \bibinfo{author}{\bibfnamefont{S.}~\bibnamefont{Tewari}},
  \bibinfo{author}{\bibfnamefont{J.~D.} \bibnamefont{Sau}}, \bibnamefont{and}
  \bibinfo{author}{\bibfnamefont{S.~D.} \bibnamefont{Sarma}},
  \bibinfo{journal}{Phys. Rev. Lett.} \textbf{\bibinfo{volume}{109}},
  \bibinfo{pages}{266402} (\bibinfo{year}{2012}).

\bibitem[{\citenamefont{Takei et~al.}(2013)\citenamefont{Takei, Fregoso, Hui,
  Lobos, and Sarma}}]{prl110.186803(2013)}
\bibinfo{author}{\bibfnamefont{S.}~\bibnamefont{Takei}},
  \bibinfo{author}{\bibfnamefont{B.~M.} \bibnamefont{Fregoso}},
  \bibinfo{author}{\bibfnamefont{H.-Y.} \bibnamefont{Hui}},
  \bibinfo{author}{\bibfnamefont{A.~M.} \bibnamefont{Lobos}}, \bibnamefont{and}
  \bibinfo{author}{\bibfnamefont{S.~D.} \bibnamefont{Sarma}},
  \bibinfo{journal}{Phys. Rev. Lett.} \textbf{\bibinfo{volume}{110}},
  \bibinfo{pages}{186803} (\bibinfo{year}{2013}).

\bibitem[{\citenamefont{Zareyan et~al.}(2002)\citenamefont{Zareyan, Belzig, and
  Nazarov}}]{prb65.184505(2002)}
\bibinfo{author}{\bibfnamefont{M.}~\bibnamefont{Zareyan}},
  \bibinfo{author}{\bibfnamefont{W.}~\bibnamefont{Belzig}}, \bibnamefont{and}
  \bibinfo{author}{\bibfnamefont{Y.~V.} \bibnamefont{Nazarov}},
  \bibinfo{journal}{Phys. Rev. B} \textbf{\bibinfo{volume}{65}},
  \bibinfo{pages}{184505} (\bibinfo{year}{2002}).

\bibitem[{\citenamefont{Sasaki et~al.}(2000)\citenamefont{Sasaki,
  De~Franceschi, Elzerman, Van~der Wiel, Eto, Tarucha, and
  Kouwenhoven}}]{nat405.764(2000)}
\bibinfo{author}{\bibfnamefont{S.}~\bibnamefont{Sasaki}},
  \bibinfo{author}{\bibfnamefont{S.}~\bibnamefont{De~Franceschi}},
  \bibinfo{author}{\bibfnamefont{J.}~\bibnamefont{Elzerman}},
  \bibinfo{author}{\bibfnamefont{W.}~\bibnamefont{Van~der Wiel}},
  \bibinfo{author}{\bibfnamefont{M.}~\bibnamefont{Eto}},
  \bibinfo{author}{\bibfnamefont{S.}~\bibnamefont{Tarucha}}, \bibnamefont{and}
  \bibinfo{author}{\bibfnamefont{L.}~\bibnamefont{Kouwenhoven}},
  \bibinfo{journal}{Nature} \textbf{\bibinfo{volume}{405}},
  \bibinfo{pages}{764} (\bibinfo{year}{2000}).

\bibitem[{\citenamefont{Thomale et~al.}(2013)\citenamefont{Thomale, Rachel, and
  Schmitteckert}}]{prb88.161103(R)(2013)}
\bibinfo{author}{\bibfnamefont{R.}~\bibnamefont{Thomale}},
  \bibinfo{author}{\bibfnamefont{S.}~\bibnamefont{Rachel}}, \bibnamefont{and}
  \bibinfo{author}{\bibfnamefont{P.}~\bibnamefont{Schmitteckert}},
  \bibinfo{journal}{Phys. Rev. B} \textbf{\bibinfo{volume}{88}},
  \bibinfo{pages}{161103} (\bibinfo{year}{2013}).

\bibitem[{\citenamefont{Gangadharaiah et~al.}(2011)\citenamefont{Gangadharaiah,
  Braunecker, Simon, and Loss}}]{prl107.036801(2011)}
\bibinfo{author}{\bibfnamefont{S.}~\bibnamefont{Gangadharaiah}},
  \bibinfo{author}{\bibfnamefont{B.}~\bibnamefont{Braunecker}},
  \bibinfo{author}{\bibfnamefont{P.}~\bibnamefont{Simon}}, \bibnamefont{and}
  \bibinfo{author}{\bibfnamefont{D.}~\bibnamefont{Loss}},
  \bibinfo{journal}{Phys. Rev. Lett.} \textbf{\bibinfo{volume}{107}},
  \bibinfo{pages}{036801} (\bibinfo{year}{2011}).

\bibitem[{\citenamefont{Sela et~al.}(2011)\citenamefont{Sela, Altland, and
  Rosch}}]{prb84.085114(2011)}
\bibinfo{author}{\bibfnamefont{E.}~\bibnamefont{Sela}},
  \bibinfo{author}{\bibfnamefont{A.}~\bibnamefont{Altland}}, \bibnamefont{and}
  \bibinfo{author}{\bibfnamefont{A.}~\bibnamefont{Rosch}},
  \bibinfo{journal}{Phys. Rev. B} \textbf{\bibinfo{volume}{84}},
  \bibinfo{pages}{085114} (\bibinfo{year}{2011}).

\bibitem[{\citenamefont{Hassler and Schuricht}(2012)}]{njp14.125018(2012)}
\bibinfo{author}{\bibfnamefont{F.}~\bibnamefont{Hassler}} \bibnamefont{and}
  \bibinfo{author}{\bibfnamefont{D.}~\bibnamefont{Schuricht}},
  \bibinfo{journal}{New J. Phys.} \textbf{\bibinfo{volume}{14}},
  \bibinfo{pages}{125018} (\bibinfo{year}{2012}).

\bibitem[{\citenamefont{Fidkowski et~al.}(2012)\citenamefont{Fidkowski, Alicea,
  Lindner, Lutchyn, and Fisher}}]{prb85.245121(2012)}
\bibinfo{author}{\bibfnamefont{L.}~\bibnamefont{Fidkowski}},
  \bibinfo{author}{\bibfnamefont{J.}~\bibnamefont{Alicea}},
  \bibinfo{author}{\bibfnamefont{N.~H.} \bibnamefont{Lindner}},
  \bibinfo{author}{\bibfnamefont{R.~M.} \bibnamefont{Lutchyn}},
  \bibnamefont{and} \bibinfo{author}{\bibfnamefont{M.~P.}
  \bibnamefont{Fisher}}, \bibinfo{journal}{Phys. Rev. B}
  \textbf{\bibinfo{volume}{85}}, \bibinfo{pages}{245121}
  (\bibinfo{year}{2012}).

\bibitem[{\citenamefont{Stoudenmire et~al.}(2011)\citenamefont{Stoudenmire,
  Alicea, Starykh, and Fisher}}]{prb84.014503(2011)}
\bibinfo{author}{\bibfnamefont{E.~M.} \bibnamefont{Stoudenmire}},
  \bibinfo{author}{\bibfnamefont{J.}~\bibnamefont{Alicea}},
  \bibinfo{author}{\bibfnamefont{O.~A.} \bibnamefont{Starykh}},
  \bibnamefont{and} \bibinfo{author}{\bibfnamefont{M.~P.}
  \bibnamefont{Fisher}}, \bibinfo{journal}{Phys. Rev. B}
  \textbf{\bibinfo{volume}{84}}, \bibinfo{pages}{014503}
  (\bibinfo{year}{2011}).

\bibitem[{\citenamefont{Potter and Lee}(2010)}]{prl105.227003(2010)}
\bibinfo{author}{\bibfnamefont{A.~C.} \bibnamefont{Potter}} \bibnamefont{and}
  \bibinfo{author}{\bibfnamefont{P.~A.} \bibnamefont{Lee}},
  \bibinfo{journal}{Phys. Rev. Lett.} \textbf{\bibinfo{volume}{105}},
  \bibinfo{pages}{227003} (\bibinfo{year}{2010}).

\bibitem[{\citenamefont{Lutchyn and Fisher}(2011)}]{prb84.214528(2011)}
\bibinfo{author}{\bibfnamefont{R.~M.} \bibnamefont{Lutchyn}} \bibnamefont{and}
  \bibinfo{author}{\bibfnamefont{M.~P.~A.} \bibnamefont{Fisher}},
  \bibinfo{journal}{Phys. Rev. B} \textbf{\bibinfo{volume}{84}},
  \bibinfo{pages}{214528} (\bibinfo{year}{2011}).

\bibitem[{\citenamefont{Wimmer et~al.}(2010)\citenamefont{Wimmer, Akhmerov,
  Medvedyeva, Tworzyd{\l}o, and Beenakker}}]{prl105.046803(2010)}
\bibinfo{author}{\bibfnamefont{M.}~\bibnamefont{Wimmer}},
  \bibinfo{author}{\bibfnamefont{A.}~\bibnamefont{Akhmerov}},
  \bibinfo{author}{\bibfnamefont{M.}~\bibnamefont{Medvedyeva}},
  \bibinfo{author}{\bibfnamefont{J.}~\bibnamefont{Tworzyd{\l}o}},
  \bibnamefont{and}
  \bibinfo{author}{\bibfnamefont{C.}~\bibnamefont{Beenakker}},
  \bibinfo{journal}{Phys. Rev. Lett.} \textbf{\bibinfo{volume}{105}},
  \bibinfo{pages}{046803} (\bibinfo{year}{2010}).

\bibitem[{\citenamefont{Potter and Lee}(2011)}]{prb83.094525(2011)}
\bibinfo{author}{\bibfnamefont{A.~C.} \bibnamefont{Potter}} \bibnamefont{and}
  \bibinfo{author}{\bibfnamefont{P.~A.} \bibnamefont{Lee}},
  \bibinfo{journal}{Phys. Rev. B} \textbf{\bibinfo{volume}{83}},
  \bibinfo{pages}{094525} (\bibinfo{year}{2011}).

\bibitem[{\citenamefont{Lutchyn et~al.}(2011)\citenamefont{Lutchyn, Stanescu,
  and Sarma}}]{prl106.127001(2011)}
\bibinfo{author}{\bibfnamefont{R.~M.} \bibnamefont{Lutchyn}},
  \bibinfo{author}{\bibfnamefont{T.~D.} \bibnamefont{Stanescu}},
  \bibnamefont{and} \bibinfo{author}{\bibfnamefont{S.~D.} \bibnamefont{Sarma}},
  \bibinfo{journal}{Phys. Rev. Lett.} \textbf{\bibinfo{volume}{106}},
  \bibinfo{pages}{127001} (\bibinfo{year}{2011}).

\bibitem[{\citenamefont{Lobos et~al.}(2012)\citenamefont{Lobos, Lutchyn, and
  Das~Sarma}}]{prl109.146403(2012)}
\bibinfo{author}{\bibfnamefont{A.~M.} \bibnamefont{Lobos}},
  \bibinfo{author}{\bibfnamefont{R.~M.} \bibnamefont{Lutchyn}},
  \bibnamefont{and}
  \bibinfo{author}{\bibfnamefont{S.}~\bibnamefont{Das~Sarma}},
  \bibinfo{journal}{Phys. Rev. Lett.} \textbf{\bibinfo{volume}{109}},
  \bibinfo{pages}{146403} (\bibinfo{year}{2012}).

\bibitem[{\citenamefont{Manolescu et~al.}(2014)\citenamefont{Manolescu,
  Marinescu, and Stanescu}}]{jpc26.172203(2014)}
\bibinfo{author}{\bibfnamefont{A.}~\bibnamefont{Manolescu}},
  \bibinfo{author}{\bibfnamefont{D.}~\bibnamefont{Marinescu}},
  \bibnamefont{and} \bibinfo{author}{\bibfnamefont{T.~D.}
  \bibnamefont{Stanescu}}, \bibinfo{journal}{J. Phys.: Cond. Matt.}
  \textbf{\bibinfo{volume}{26}}, \bibinfo{pages}{172203}
  (\bibinfo{year}{2014}).

\bibitem[{\citenamefont{Cayao et~al.}(2015)\citenamefont{Cayao, Prada,
  San-Jose, and Aguado}}]{prb91.024514(2015)}
\bibinfo{author}{\bibfnamefont{J.}~\bibnamefont{Cayao}},
  \bibinfo{author}{\bibfnamefont{E.}~\bibnamefont{Prada}},
  \bibinfo{author}{\bibfnamefont{P.}~\bibnamefont{San-Jose}}, \bibnamefont{and}
  \bibinfo{author}{\bibfnamefont{R.}~\bibnamefont{Aguado}},
  \bibinfo{journal}{Phys. Rev. B} \textbf{\bibinfo{volume}{91}},
  \bibinfo{pages}{024514} (\bibinfo{year}{2015}).

\bibitem[{\citenamefont{Setiawan et~al.}(2015)\citenamefont{Setiawan, Brydon,
  Sau, and Sarma}}]{prb91.2145413(2015)}
\bibinfo{author}{\bibfnamefont{F.}~\bibnamefont{Setiawan}},
  \bibinfo{author}{\bibfnamefont{P.}~\bibnamefont{Brydon}},
  \bibinfo{author}{\bibfnamefont{J.~D.} \bibnamefont{Sau}}, \bibnamefont{and}
  \bibinfo{author}{\bibfnamefont{S.~D.} \bibnamefont{Sarma}},
  \bibinfo{journal}{Phys. Rev. B} \textbf{\bibinfo{volume}{91}},
  \bibinfo{pages}{214513} (\bibinfo{year}{2015}).

\bibitem[{\citenamefont{Liu et~al.}(2013)\citenamefont{Liu, Zhang, and
  Law}}]{prb88.064509(2013)}
\bibinfo{author}{\bibfnamefont{J.}~\bibnamefont{Liu}},
  \bibinfo{author}{\bibfnamefont{F.-C.} \bibnamefont{Zhang}}, \bibnamefont{and}
  \bibinfo{author}{\bibfnamefont{K.~T.} \bibnamefont{Law}},
  \bibinfo{journal}{Phys. Rev. B} \textbf{\bibinfo{volume}{88}},
  \bibinfo{pages}{064509} (\bibinfo{year}{2013}).

\bibitem[{\citenamefont{San-Jose et~al.}(2013)\citenamefont{San-Jose, Cayao,
  Prada, and Aguado}}]{njp15.075019(2013)}
\bibinfo{author}{\bibfnamefont{P.}~\bibnamefont{San-Jose}},
  \bibinfo{author}{\bibfnamefont{J.}~\bibnamefont{Cayao}},
  \bibinfo{author}{\bibfnamefont{E.}~\bibnamefont{Prada}}, \bibnamefont{and}
  \bibinfo{author}{\bibfnamefont{R.}~\bibnamefont{Aguado}},
  \bibinfo{journal}{New J. of Phys.} \textbf{\bibinfo{volume}{15}},
  \bibinfo{pages}{075019} (\bibinfo{year}{2013}).

\bibitem[{\citenamefont{Komijani and Affleck}(2014)}]{prb90.115107(2014)}
\bibinfo{author}{\bibfnamefont{Y.}~\bibnamefont{Komijani}} \bibnamefont{and}
  \bibinfo{author}{\bibfnamefont{I.}~\bibnamefont{Affleck}},
  \bibinfo{journal}{Physical Review B} \textbf{\bibinfo{volume}{90}},
  \bibinfo{pages}{115107} (\bibinfo{year}{2014}).

\bibitem[{\citenamefont{Ihnatsenka and Zozoulenko}(2006)}]{prb73.075331(2006)}
\bibinfo{author}{\bibfnamefont{S.}~\bibnamefont{Ihnatsenka}} \bibnamefont{and}
  \bibinfo{author}{\bibfnamefont{I.}~\bibnamefont{Zozoulenko}},
  \bibinfo{journal}{Phys. Rev. B} \textbf{\bibinfo{volume}{73}},
  \bibinfo{pages}{075331} (\bibinfo{year}{2006}).

\bibitem[{\citenamefont{Yu et~al.}(2011)\citenamefont{Yu, Qi, Bernevig, Fang,
  and Dai}}]{prb84.075119(2011)}
\bibinfo{author}{\bibfnamefont{R.}~\bibnamefont{Yu}},
  \bibinfo{author}{\bibfnamefont{X.~L.} \bibnamefont{Qi}},
  \bibinfo{author}{\bibfnamefont{A.}~\bibnamefont{Bernevig}},
  \bibinfo{author}{\bibfnamefont{Z.}~\bibnamefont{Fang}}, \bibnamefont{and}
  \bibinfo{author}{\bibfnamefont{X.}~\bibnamefont{Dai}},
  \bibinfo{journal}{Phys. Rev. B} \textbf{\bibinfo{volume}{84}},
  \bibinfo{pages}{075119} (\bibinfo{year}{2011}).

\bibitem[{\citenamefont{Xu et~al.}(2008)\citenamefont{Xu, Heinzel, Evaldsson,
  and Zozoulenko}}]{prb77.245401(2008)}
\bibinfo{author}{\bibfnamefont{H.}~\bibnamefont{Xu}},
  \bibinfo{author}{\bibfnamefont{T.}~\bibnamefont{Heinzel}},
  \bibinfo{author}{\bibfnamefont{M.}~\bibnamefont{Evaldsson}},
  \bibnamefont{and}
  \bibinfo{author}{\bibfnamefont{I.}~\bibnamefont{Zozoulenko}},
  \bibinfo{journal}{Phys. Rev. B} \textbf{\bibinfo{volume}{77}},
  \bibinfo{pages}{245401} (\bibinfo{year}{2008}).

\bibitem[{\citenamefont{Xu and Heinzel}(2013)}]{pla377.3148(2013)}
\bibinfo{author}{\bibfnamefont{H.}~\bibnamefont{Xu}} \bibnamefont{and}
  \bibinfo{author}{\bibfnamefont{T.}~\bibnamefont{Heinzel}},
  \bibinfo{journal}{Phys. Lett. A} \textbf{\bibinfo{volume}{377}},
  \bibinfo{pages}{3148} (\bibinfo{year}{2013}).

\bibitem[{\citenamefont{Tewari et~al.}(2012)\citenamefont{Tewari, Stanescu,
  Sau, and Sarma}}]{prb86.024505(2012)}
\bibinfo{author}{\bibfnamefont{S.}~\bibnamefont{Tewari}},
  \bibinfo{author}{\bibfnamefont{T.}~\bibnamefont{Stanescu}},
  \bibinfo{author}{\bibfnamefont{J.~D.} \bibnamefont{Sau}}, \bibnamefont{and}
  \bibinfo{author}{\bibfnamefont{S.~D.} \bibnamefont{Sarma}},
  \bibinfo{journal}{Phys. Rev. B} \textbf{\bibinfo{volume}{86}},
  \bibinfo{pages}{024504} (\bibinfo{year}{2012}).

\bibitem[{\citenamefont{San-Jose et~al.}(2015)\citenamefont{San-Jose, Lado,
  Aguado, Guinea, and Fern{\'a}ndez-Rossier}}]{prx5.041042(2015)}
\bibinfo{author}{\bibfnamefont{P.}~\bibnamefont{San-Jose}},
  \bibinfo{author}{\bibfnamefont{J.}~\bibnamefont{Lado}},
  \bibinfo{author}{\bibfnamefont{R.}~\bibnamefont{Aguado}},
  \bibinfo{author}{\bibfnamefont{F.}~\bibnamefont{Guinea}}, \bibnamefont{and}
  \bibinfo{author}{\bibfnamefont{J.}~\bibnamefont{Fern{\'a}ndez-Rossier}},
  \bibinfo{journal}{Phys. Rev. X} \textbf{\bibinfo{volume}{5}},
  \bibinfo{pages}{041042} (\bibinfo{year}{2015}).

\bibitem[{\citenamefont{Xu et~al.}(2010)\citenamefont{Xu, Heinzel, Shylau, and
  Zozoulenko}}]{prb82.115311(2010)}
\bibinfo{author}{\bibfnamefont{H.}~\bibnamefont{Xu}},
  \bibinfo{author}{\bibfnamefont{T.}~\bibnamefont{Heinzel}},
  \bibinfo{author}{\bibfnamefont{A.~A.} \bibnamefont{Shylau}},
  \bibnamefont{and}
  \bibinfo{author}{\bibfnamefont{I.}~\bibnamefont{Zozoulenko}},
  \bibinfo{journal}{Phys. Rev. B} \textbf{\bibinfo{volume}{82}},
  \bibinfo{pages}{115311} (\bibinfo{year}{2010}).

\bibitem[{\citenamefont{Avetisyan et~al.}(2009)\citenamefont{Avetisyan,
  Partoens, and Peeters}}]{prb79.035421(2009)}
\bibinfo{author}{\bibfnamefont{A.}~\bibnamefont{Avetisyan}},
  \bibinfo{author}{\bibfnamefont{B.}~\bibnamefont{Partoens}}, \bibnamefont{and}
  \bibinfo{author}{\bibfnamefont{F.}~\bibnamefont{Peeters}},
  \bibinfo{journal}{Phys. Rev. B} \textbf{\bibinfo{volume}{79}},
  \bibinfo{pages}{035421} (\bibinfo{year}{2009}).

\bibitem[{\citenamefont{Prada et~al.}(2012)\citenamefont{Prada, San-Jose, and
  Aguado}}]{prb86.180503(R)(2012)}
\bibinfo{author}{\bibfnamefont{E.}~\bibnamefont{Prada}},
  \bibinfo{author}{\bibfnamefont{P.}~\bibnamefont{San-Jose}}, \bibnamefont{and}
  \bibinfo{author}{\bibfnamefont{R.}~\bibnamefont{Aguado}},
  \bibinfo{journal}{Phys. Rev. B} \textbf{\bibinfo{volume}{86}},
  \bibinfo{pages}{180503} (\bibinfo{year}{2012}).

\end{thebibliography}
\end{document}